\documentclass[prb,preprint,amsmath,amssymb,showpacs,floatfix,tightenlines]{revtex4}
\usepackage{graphicx}
\usepackage{dcolumn}
\usepackage{bm}

\newcommand{\cm}{cm$^{-1}$}

\begin{document}
\title{Symmetry disquisition on the TiOX phase diagram}
\author{Daniele Fausti}
\email{d.fausti@rug.nl}
\author{Tom T. A. Lummen}
\author{Cosmina Angelescu}
\author{Roberto Macovez}
\author{Javier Luzon}
\author{Ria Broer}
\author{Petra Rudolf}
\author{Paul H.M. van Loosdrecht}
\email{P.H.M.van.Loosdrecht@rug.nl} \affiliation{Zernike Institute
for Advanced Materials, University of Groningen, 9747 AG Groningen,
The Netherlands.}
\author{Natalia Tristan}
\author{Bernd B\"uchner}
\affiliation{IFW Dresden, D-01171 Dresden, Germany }
\author{Sander van Smaalen}
\affiliation{Laboratory of Crystallography, University of
Bayreuth, 95440 Bayreuth, Germany}
\author{Angela M\"oller}
\author{Gerd Meyer}
\author{Timo Taetz}
\affiliation{Institut f\"ur Anorganische Chemie, Universit\"at zu
K\"oln, 50937 K\"oln, Germany}
\date{\today}
\pacs{71.20;73.43;78.30;63.2;68.35}

\begin{abstract}
The sequence of phase transitions and the symmetry of in particular
the low temperature incommensurate and spin-Peierls phases of the
quasi one-dimensional inorganic spin-Peierls system TiOX (TiOBr and
TiOCl) have been studied using inelastic light scattering
experiments. The anomalous first-order character of the transition
to the spin-Peierls phase is found to be a consequence of the
different symmetries of the incommensurate and spin-Peierls
(P$2_{1}/m$) phases.

The pressure dependence of the lowest transition temperature
strongly suggests that magnetic interchain interactions play an
important role in the formation of the spin-Peierls and the
incommensurate phases. Finally, a comparison of Raman data on VOCl
to the TiOX spectra shows that the high energy scattering observed
previously has a phononic origin.
\end{abstract}

\pacs{
68.18.Jk Phase transitions\\
63.20.-e Phonons in crystal lattices\\
75.30.Et Exchange and superexchange interactions\\
75.30.Kz Magnetic phase boundaries (including magnetic transitions, metamagnetism, etc.)\\
78.30.-j Infrared and Raman spectra
}

\maketitle

\section{Introduction}
The properties of low-dimensional spin systems are one of the key
topics of contemporary condensed matter physics. Above all, the
transition metal oxides with highly anisotropic interactions and
low-dimensional structural elements provide a fascinating playground
to study novel phenomena, arising from their low-dimensional nature
and from the interplay between lattice, orbital, spin and charge
degrees of freedom. In particular, low-dimensional quantum spin
(S=1/2) systems have been widely discussed in recent years. Among
them, layered systems based on a $3d^{9}$\ electronic configuration
were extensively studied in view of the possible relevance of
quantum magnetism to high temperature
superconductivity\cite{Imad98,Dag99}. Though they received less
attention, also spin=1/2 systems based on early transition metal
oxides with electronic configuration $3d^{1}$, such as titanium
oxyhalides (TiOX, with X=Br or Cl), exhibit a variety of interesting
properties\cite{kataev2003,imai2003}. The attention originally
devoted to the layered quasi two-dimensional $3d^{1}$\
antiferromagnets arose from considering them as the electron analog
to the high-$T_{c}$\ cuprates\cite{Maule88}. Only recently TiOX
emerged in a totally new light, namely as a one-dimensional
antiferromagnet and as the second example of an inorganic
spin-Peierls compound (the first being
CuGeO$_{3}$)\cite{seidel2003,caimi2004}.

The TiO bilayers constituting the TiOX lattice are candidates for
various exotic electronic configurations, such as orbital
ordered\cite{kataev2003}, spin-Peierls\cite{seidel2003} and
resonating-valence-bond states\cite{Beynon1993}. In the case of the
TiOX family the degeneracy of the $d$\ orbitals is completely
removed by the crystal field splitting, so that the only
$d-$electron present, mainly localized on the Ti site, occupies a
nondegenerate energy orbital\cite{kataev2003}. As a consequence of
the shape of the occupied orbital (which has lobes oriented in the
$b-$ and $c-$directions, where $c$\ is perpendicular to the layers),
the exchange interaction between the spins on different Ti ions
arises mainly from direct exchange within the TiO bilayers, along
the $b$\ crystallographic direction\cite{kataev2003}. This, in spite
of the two-dimensional structural character, gives the magnetic
system of the TiOX family its peculiar quasi one-dimensional
properties\cite{seidel2003}. Magnetic
susceptibility\cite{seidel2003} and ESR \cite{kataev2003}
measurements at high temperature are in reasonably good agreement
with an antiferromagnetic, one-dimensional spin-1/2 Heisenberg chain
model. At low temperature ($T_{c1}$) TiOX shows a first-order phase
transition to a dimerised nonmagnetic state, discussed in terms of a
spin Peierls state \cite{seidel2003,caimi2004-1,shaz}. Between this
low temperature spin Peierls phase (SP) and the one-dimensional
antiferromagnet in the high temperature phase (HT), various
experimental evidence \cite{hemberger,ruck2005,imai2003,Lemmens2003}
showed the existence of an intermediate phase, whose nature and
origin is still debated. The temperature region of the intermediate
phase is different for the two compounds considered in this work,
for TiOBr $T_{c1}=28$~K and $T_{c2}=48$~K while for TiOCl
$T_{c1}=67$~K and $T_{c2}=91$~K. To summarize the properties so far
reported, the intermediate phase ($T_{c1}<T_{c2}$) exhibits a gapped
magnetic excitation spectrum\cite{imai2003}, anomalous broadening of
the phonon modes in Raman and IR
spectra\cite{Lemmens2003,caimi2004-1}, and features of a periodicity
incommensurate with the
lattice\cite{palatinus2005,smaalen2005,Schon2006,krim2006}.
Moreover, the presence of a pressure induced metal to insulator
transition has been recently suggested for TiOCl\cite{kun2006}. Due
to this complex phase behavior, both TiOCl and TiOBr have been
extensively discussed in recent literature, and various questions
still remain open: there is no agreement on the crystal symmetry of
the spin Peierls phase, the nature and symmetry of the
incommensurate phase is not clear and the anomalous first-order
character of the transition to the spin Peierls state is not
explained.

Optical methods like Raman spectroscopy are powerful experimental
tools for revealing the characteristic energy scales associated with
the development of broken symmetry ground states, driven by magnetic
and structural phase transitions. Indeed, information on the nature
of the magnetic ground state, lattice distortion, and interplay of
magnetic and lattice degrees of freedom can be obtained by studying
in detail the magnetic excitations and the phonon spectrum as a
function of temperature. The present paper reports on a vibrational
Raman study of TiOCl and TiOBr, a study of the symmetry properties
of the three phases and gives coherent view of the anomalous first
order character of the transition to the spin Peierls phase. Through
pressure-dependence measurements of the magnetic susceptibility, the
role of magnon-phonon coupling in determining the complex phase
diagram of TiOX is discussed. Finally, via a comparison with the
isostructural compound VOCl, the previously
reported\cite{Lemmens2003,lemmens2005} high energy scattering is
revisited, ruling out a possible interpretation in terms of magnon
excitations.

\section{Experiment}
Single crystals of TiOCl, TiOBr, and VOCl have been grown by a
chemical vapor transport technique. The crystallinity was checked by
X-ray diffraction\cite{ruck2005}. Typical crystal dimensions are a
few mm$^{2}$\ in the $ab-$plane and 10-100~$\mu$m along the
$c-$axis, the stacking direction\cite{smaalen2005}. The sample was
mounted in an optical flow cryostat, with a temperature
stabilization better than 0.1 K in the range from 2.6 K to 300 K.
The Raman measurements were performed using a triple grating
micro-Raman spectrometer (Jobin Yvon, T64000), equipped with a
liquid nitrogen cooled CCD detector (resolution 2~\cm\ for the
considered frequency interval). The experiments were performed with
a 532 nm Nd:YVO$_{4}$\ laser. The power density on the sample was
kept below 500 W/cm$^{2}$\ to avoid sample degradation and to
minimize heating effects.

The polarization was controlled on both the incoming and outgoing
beam, giving access to all the polarizations schemes allowed by the
back-scattering configuration. Due to the macroscopic morphology of
the samples (thin sheets with natural surfaces parallel to the
$ab-$planes) the polarization analysis was performed mainly with the
incoming beam parallel to the $c-$axis ($c$(aa)$\bar{c}$,
$c$(ab)$\bar{c}$\ and $c$(bb)$\bar{c}$, in Porto notation). Some
measurements were performed with the incoming light polarized along
the $c-$axis, where the $k-$vector of the light was parallel to the
$ab-$plane and the polarization of the outgoing light was not
controlled. These measurements will be labeled as
$x$($c\star$)$\bar{x}$.

The magnetization measurements were performed in a
Quantum Design Magnetic Property Measurement System. The
pressure cell used is specifically designed for measurement of the
DC-magnetization in order to minimize the cell's magnetic
response. The cell was calibrated using the lead superconducting
transition as a reference, and the cell's signal (measured at
atmospheric pressure) was subtracted from the data.

\section{Results and Discussion}
The discussion will start with a comparison of Raman experiments on
TiOCl and TiOBr in the high temperature phase, showing the
consistency with the reported structure. Afterwards, through the
analysis of Raman spectra the crystal symmetry in the low
temperature phases will be discussed, and in the final part a
comparison with the isostructural VOCl will be helpful to shed some
light on the origin of the anomalous high energy scattering reported
for TiOCl and TiOBr\cite{Lemmens2003,lemmens2005}.

\subsection{High Temperature Phase}

The crystal structure of TiOX in the high temperature (HT) phase
consists of buckled Ti-O bilayers separated by layers of X ions. The
HT structure is orthorhombic with space group P$mmn$. The full
representation\cite{Rou1981} of the vibrational modes in this space
group is:
\begin{equation}
\Gamma_{tot}=3A_{g}+2B_{1u}+3B_{2g}+2B_{2u}+3B_{3g}+2B_{3u}.
\end{equation}
Among these, the modes with symmetry $B_{1u}$, $B_{2u}$, and
$B_{3u}$ are infrared active in the polarizations along the $c$,
$b$, and $a$\ crystallographic axes\cite{caimi2004-1}, respectively.
The modes with symmetry $A_{g}$, $B_{2g}$, and $B_{3g}$\ are
expected to be Raman active: The $A_{g}$\ modes in the polarization
($aa$), ($bb$), and ($cc$); the $B_{2g}$\ modes in ($ac$) and the
$B_{3g}$ ones in ($bc$).
\begin{figure}[htb]
 \centerline{\includegraphics[width=70mm]{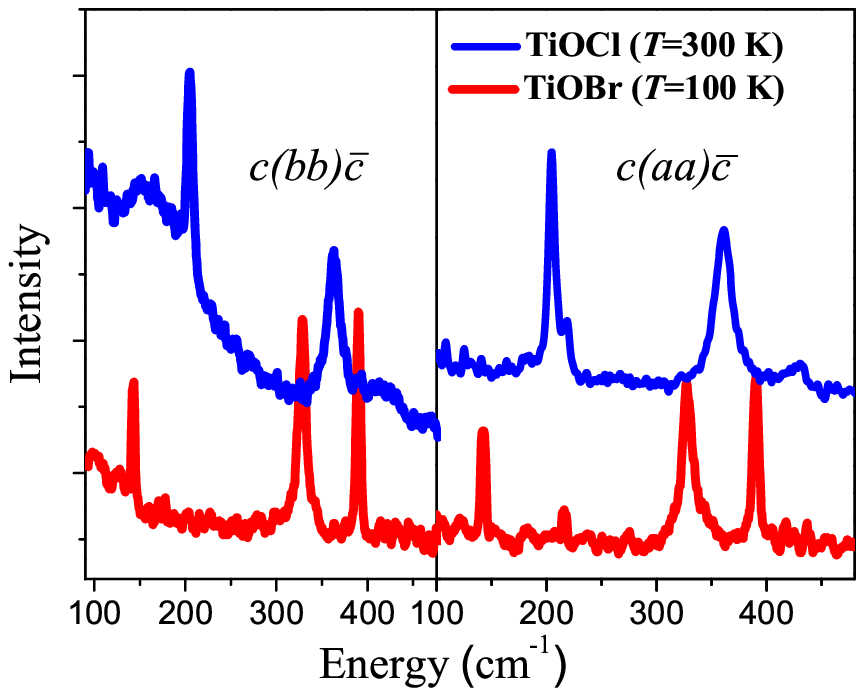}}
 \caption{(Color online) Polarized Raman spectra ($A_{g}$) of TiOCl and TiOBr
           in the high temperature phase, showing the three $A_{g}$ modes. Left panel: ($bb$) polarization;
           right panel: ($aa$) polarization.}
 \label{fig1plus}
\end{figure}
Fig.\ref{fig1plus} shows the room temperature Raman
measurements in different polarizations for TiOCl and TiOBr, and
\begin{figure}[htb]
 \centerline{\includegraphics[width=90mm]{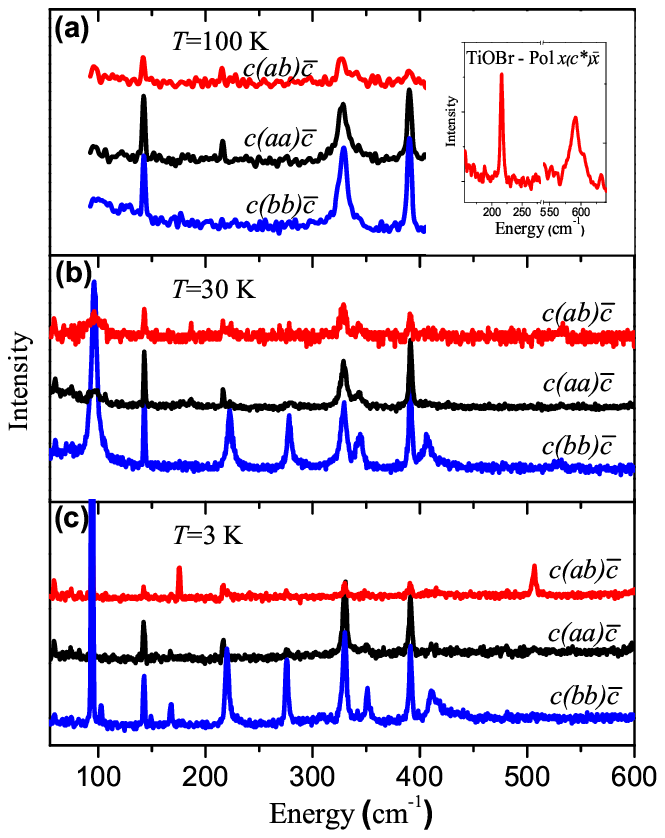}}
 \caption{(Color online)
 Polarization analysis of the Raman spectra in the three phases of TiOBr,
 taken at 3 (a), 30 (b) and
 100K (c). The spectra of TiOCl show the same main features and closely
resemble those of TiOBr. Table \ref{Tab2} reports the frequencies of the
TiOCl modes. The inset shows the TiOBr spectrum in the $x$($c\ast$)$\bar{x}$
polarization (see text).}
 \label{fig1}
\end{figure}
Fig.\ref{fig1} displays the characteristic Raman spectra for the
three different phases of TiOBr, the spectra are taken at 100 (a),
30 (b) and 3K (c). At room temperature three Raman active modes are
clearly observed in both compounds for the $c$($aa$)$\bar{c}$\ and
$c$($bb$)$\bar{c}$\ polarizations (Fig.\ref{fig1plus}), while none
are observed in the $c$($ab$)$\bar{c}$\ polarization. These results
are in good agreement with the group theoretical analysis. The
additional weakly active modes observed at 219~\cm\ for TiOCl and at
217~\cm\ for TiOBr are ascribed to a leak from a different
polarization. This is confirmed by the measurements with the optical
axis parallel to the $ab$-planes ($x$($c\star$)$\bar{x}$) on TiOBr,
where an intense mode is observed at the same frequency (as shown in
the inset of Fig.\ref{fig1}(a)). In addition to these expected
modes, TiOCl displays a broad peak in the $c$($bb$)$\bar{c}$
polarization, centered at around 160~\cm\ at 300K; a similar feature
is observed in TiOBr as a broad background in the low frequency
region at 100K. As discussed for TiOCl\cite{Lemmens2003}, these
modes are thought to be due to pre-transitional fluctuations. Upon
decreasing the temperature, this "peaked" background first softens,
resulting in a broad mode at $T_{c2}$\ (see Fig.\ref{fig1}(b)), and
then locks at $T_{c1}$ into an intense sharp mode at 94.5~\cm\ for
TiOBr (Fig.\ref{fig1}(c)) and at 131.5~\cm\ for TiOCl.

\begin{table}[htb]
\caption{(a)Vibrational modes for the high temperature phase in
TiOCl, TiOBr and VOCl. The calculated values are obtained with a
spring model. The mode reported in $italics$\ in Table \ref{Tab1}
are measured in the $x$($c\star$)$\bar{x}$\ polarization they could
therefore have either $B_{2g}$\ or $B_{3g}$\ symmetry (see
experimental details).}
 \begin{ruledtabular}
 \begin{tabular}{lcccccc}
(a)&TiOBr& & TiOCl & & VOCl&\\
&Exp. & Cal.&Exp. & Cal.&Exp. & Cal.\\
\hline
$A_{g}~(\sigma_{aa},\sigma_{bb},\sigma_{cc})$ & 142.7
    & 141 & 203 & 209.1 &201 &208.8 \\
 & 329.8 & 328.2 & 364.8 & 331.2 &384.9 &321.5 \\
 & 389.9 & 403.8 & 430.9 & 405.2 &408.9 & 405.2 \\
$B_{2g}(\sigma_{ac})$ & & 105.5& & 157.1 & & 156.7 \\
 & & 328.5& & 330.5 & & 320.5 \\
 & & 478.2& & 478.2 & & 478.2 \\
$B_{3u}(IR,a)$ &77\footnotemark[1] &75.7 &104\footnotemark[2] & 94.4 & &93.7 \\
 &417\footnotemark[1] &428.5 &438\footnotemark[2] &428.5 & &425.2\\
$B_{3g}(\sigma_{bc})$ &\textit{60} &86.4 & &129.4 & &129.4\\
 &\textit{216} &336.8 & \textit{219}\footnotemark[3]&336.8& &327.2\\
 &\textit{598} &586.3 & &586.3 & &585.6\\
$B_{2u}(IR,b)$ &131\footnotemark[1] &129.1 &176\footnotemark[2] &160.8 & &159.5\\
 &275\footnotemark[1] &271.8 &294\footnotemark[2] &272.1 & &269.8\\
$B_{1u}(IR,c)$ & &155.7 & &194.1 & &192.4\\
 & &304.8 & &301.1 & &303.5\\
\end{tabular}\vspace*{5mm}
 \end{ruledtabular}
 \footnotetext[1]{Value taken from Ref.\cite{caimi2004}.}
 \footnotetext[2]{Value taken from Ref.\cite{caimi2004-1}.}
 \footnotetext[3]{Value obtained considering the leakage in
         the $\sigma_{yy}$\ polarization.}
\label{Tab1}
\end{table}

\begin{table}[htb]
\caption{The ratio between the frequency of the $A_{g}$ Raman active
modes measured in TiOBr and TiOCl is related to the atomic
displacements of the different modes as calculated for TiOBr (all
the eigenvectors are fully $c-$polarized, the values are normalized
to the largest displacement).}
 \begin{ruledtabular}
\begin{tabular}{lcccccc}
(b) & Mode &$\nu$(TiOBr) &$\nu_{Cl}/\nu_{Br}$ & Ti & O & Br \\
 &1 & 142.7 & 1.42 &0.107& 0.068&1 \\
 &2 & 329.8 & 1.11 &1 &0.003 & 0.107\\
 &3 & 389.9 & 1.11 &0.04 & 1 &0.071\\
\end{tabular}
 \end{ruledtabular}
\label{Tab1b}
\end{table}

The frequency of all the vibrational modes observed for TiOCl and
TiOBr in their high temperature phase are summarized in Table
\ref{Tab1}. Here, the infrared active modes are taken from the
literature\cite{caimi2004-1,caimi2004} and for the Raman modes the
temperatures chosen for the two compounds are 300K for TiOCl and
100K for TiOBr. The observed Raman frequencies agree well with
previous reports\cite{Lemmens2003}. The calculated values reported
in Table \ref{Tab1} are obtained with a spring-model calculation
based on phenomenological longitudinal and transversal spring
constants (see Appendix). The spring constants used were optimized
using the TiOBr experimental frequencies (except for the ones of the
$B_{3g}$\ modes due to their uncertain symmetry) and kept constant
for the other compounds. The frequencies for the other two compounds
are obtained by merely changing the appropriate atomic masses and
are in good agreement with the experimental values. The relative
atomic displacements for each mode of $A_{g}$\ symmetry are shown in
Table \ref{Tab1b}. The scaling ratio for the lowest frequency mode
(mode 1) between the two compounds is in good agreement with the
calculation of the atomic displacements. The low frequency mode is
mostly related to Br/Cl movement and, indeed, the ratio
$\nu_{TiOCl}/\nu_{TiOBr}=1.42$\ is similar to the mass ratio
$\sqrt{M_{Br}}/\sqrt{M_{Cl}}$. The other modes (2 and 3) involve
mainly Ti or O displacements, and their frequencies scale with a
lower ratio, as can be expected.

\subsection{Low Temperature Phases}
Although the symmetry of the low temperature phases has been studied
by X-ray crystallography, there is no agreement concerning the
symmetry of the SP phase; different works proposed two different
space groups, P$2_{1}/m$\cite{Schon2006,palatinus2005,smaalen2005}
and P$mm2$\cite{sasaki2006}.

The possible symmetry changes that a dimerisation of Ti ions in the
$b-$direction can cause are considered in order to track down the
space group of the TiOX crystals in the low temperature phases.
Assuming that the low temperature phases belong to a subgroup of the
high temperature orthorhombic space group P$mmn$, there are
different candidate space groups for the low temperature phases.
Note that the assumption is certainly correct for the intermediate
phase, because the transition at $T_{c2}$\ is of second-order
implying a symmetry reduction, while it is not necessarily correct
for the low temperature phase, being the transition at $T_{c1}$ is
of first-order.

\begin{figure}[htb]
 \includegraphics[width=80mm]{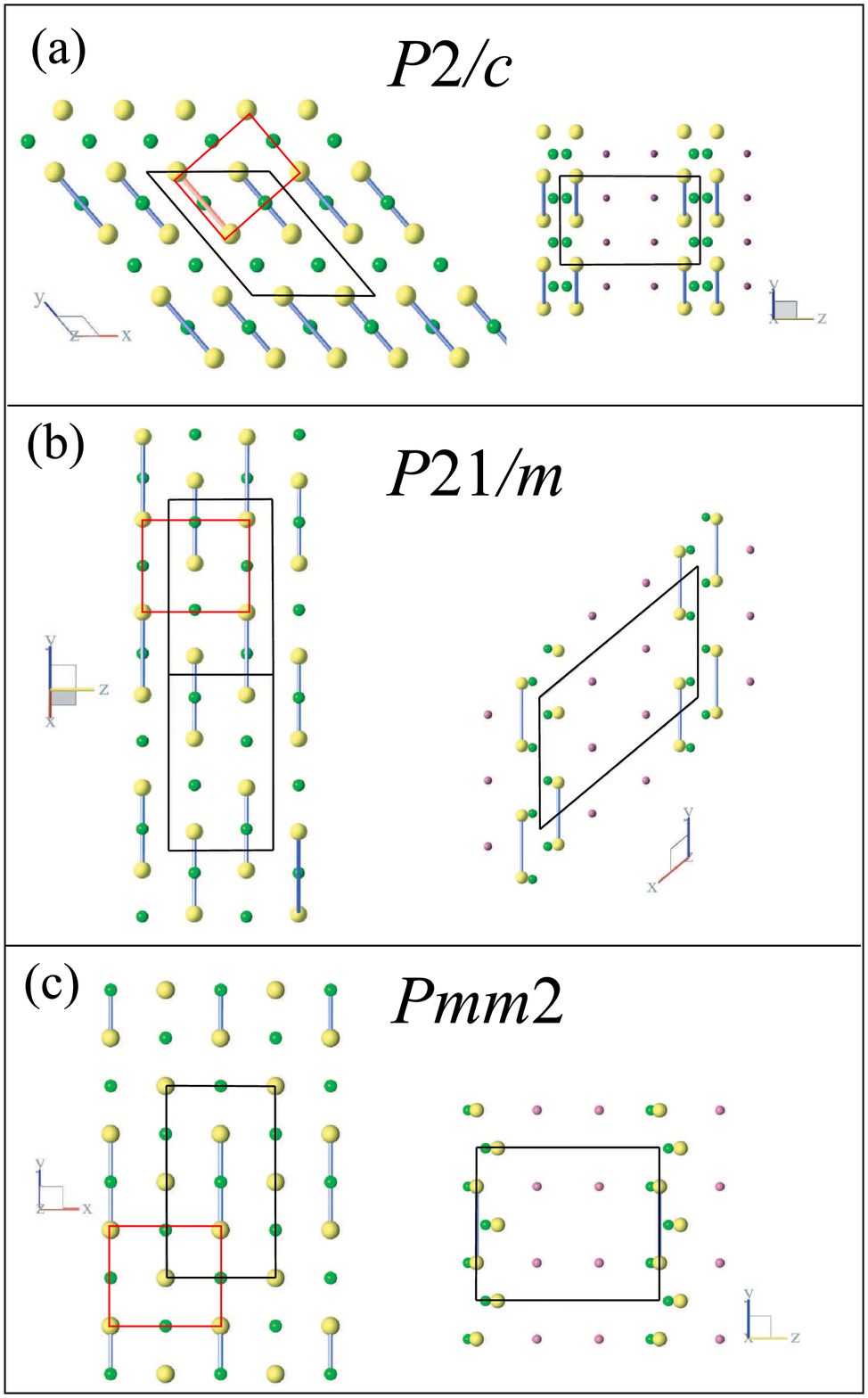}
 \caption{(Color online) Comparison of the possible low
temperature symmetries. The low temperature structures reported are
discussed, considering a dimerisation of the unit cell due to Ti-Ti
coupling and assuming a reduction of the crystal symmetry. The red
rectangle denotes the unit cell of the orthorhombic HT structure.
Structure (a) is monoclinic with its unique axis parallel to the
orthorhombic $c-$axis (space group P$2/c$), (b) shows the suggested
monoclinic structure for the SP phase (P$2_{1}/m$), and (c) depicts
the alternative orthorhombic symmetry proposed for the low T phase
P$mm2$.}
 \label{fig2}
\end{figure}
\begin{table}[htb]
 \caption{Comparison between the possible low temperature space group.}
 \label{Tab3}
 \begin{ruledtabular}
 \begin{tabular}{lc}
(a)& Space group P$2/c$ \\
 & Unique axis $\perp$\ to TiO plane, $C_{2h}^{4}$ \\
 & 4TiOBr per unit cell \\
 & $\Gamma = 7A_{g} + 6A_{u} + 9B_{g} + 11B_{u}$ \\
 & $7A_{g}$\ Raman active $\sigma_{xx},\sigma_{yy},\sigma_{zz},\sigma_{xy}$\\
 & $11B_{g}$\ Raman active $\sigma_{xz},\sigma_{yz}$ \\
 & $6A_{u}$\ and $9B_{u}$\ IR active \\
\hline
(b)& Space group P$2_{1}/m$ \\
 &Unique axis in the TiO plane, $C_{2h}^{2}$\\
 &4 TiOBr per unit cell \\
 &$\Gamma = 12A_{g} + 5A_{u} + 6B_{g} + 10B_{u}$ \\
 &$12A_{g}$\ Raman active $\sigma_{xx},\sigma_{yy},\sigma_{zz},\sigma_{xy}$\\
 &$6B_{g}$\ Raman active $\sigma_{xz},\sigma_{yz}$\\
 &$5A_{u}$\ and $10B_{u}$\ IR active \\
\hline
(c)& Space group P$mm2$\\
 &4 TiOBr per unit cell\\
 &$\Gamma = 11A_{1} + A_{2} + 4B_{1} + 5B_{2}$\\
 &$11A_{1}$\ Raman active $\sigma_{xx},\sigma_{yy},\sigma_{zz}$\\
 &$A_{2}$\ Raman active $\sigma_{xy}$\\
 &$4B_{1}$\ and $5B_{2}$\ Raman active in $\sigma_{xz}$\ and $\sigma_{yz}$
 \end{tabular}
 \end{ruledtabular}
\end{table}
Fig.\ref{fig2} shows a sketch of the three possible low temperature
symmetries considered, and Table \ref{Tab3} reports a summary of the
characteristic of the unit cell together with the number of phonons
expected to be active for the different space groups. Depending on
the relative position of the neighboring dimerised Ti pairs, the
symmetry elements lost in the dimerisation are different and the
possible space groups in the SP phase are P$2/c$\ (Table
\ref{Tab3}(a)), P$2_{1}/m$\ (b) or P$mm2$\ (c). The first two are
monoclinic groups with their unique axis perpendicular to the TiO
plane (along the $c-$axis of the orthorhombic phase), and lying in
the TiO plane ($\parallel$\ to the $a-$axis of the orthorhombic
phase), respectively. The third candidate (Fig.\ref{fig2}(c)) has
orthorhombic symmetry.

The group theory analysis based on the two space groups suggested
for the SP phase (P$2_{1}/m$\cite{palatinus2005} and
P$mm2$\cite{sasaki2006}) shows that the number of modes expected to
be Raman active is different in the two cases (Table \ref{Tab3}(b)
and (c)). In particular, the 12 fully symmetric vibrational modes
($A_{g}$), in the P$2_{1}/m$\ space group, are expected to be active
in the $\sigma_{xx},\sigma_{yy},\sigma_{zz}$ and $\sigma_{xy}$\
polarizations, and $6 B_{g}$\ modes are expected to be active in the
cross polarizations ($\sigma_{xz}$\ and $\sigma_{yz}$). Note that in
this notation, $z$ refers to the unique axis of the monoclinic cell,
so $\sigma_{yz}$\ corresponds to $c(ab)c$\ for the HT orthorhombic
phase. For P$mm2$\ the 11 $A_{1}$ vibrational modes are expected to
be active in the $\sigma_{xx},\sigma_{yy},\sigma_{zz}$\
polarizations, and only one mode of symmetry $A_{2}$\ is expected to
be active in the cross polarization ($\sigma_{xy}$\ or $c(ab)c$).
\begin{table}[htb]
 \caption{Vibrational modes of the low temperature phases.}
 \label{Tab2}
\begin{ruledtabular}
 \begin{tabular}{llcccccc}
\multicolumn{8}{c}{spin Peierls phase}\\
\hline
(a) &TiOBr&$A_{g} (\sigma_{xx},\sigma_{yy})$ & 94.5 & 102.7 & 142.4 & 167 & 219\\
 & & & 276.5 & 330 & 351 & 392 & 411$^{\ast}$\\
 & & $A_{g} (\sigma_{xy})$ & 175,6 & 506.5 \\
 &TiOCl & $A_{g} (\sigma_{xx},\sigma_{yy})$ & 131.5 & 145.8 & 203.5 & 211.5 & 296.5\\
 & & & 305.3 & 322.6 & 365.1 & 387.5 & 431$^{\ast}$\\
 & &$A_{g} (\sigma_{xy})$ & 178.5 & 524.3\\
\end{tabular}\vspace*{5mm}

\begin{tabular}{llcccccc}
\multicolumn{8}{c}{Intermediate phase} \\
\hline
(b) &TiOBr (30K)& $A_{g} (\sigma_{xx},\sigma_{yy})$ & 94.5 & 142 & 221.5 & 277 & 328.5 \\
 & & & 344.5 & 390.4 \\
 & TiOCl (75K)& $A_{g} (\sigma_{xx},\sigma_{yy})$ & 132.8 & 206.2 & 302 & 317.2 & 364.8\\
 && & 380 & 420.6
\\
\end{tabular}
\end{ruledtabular}
\footnotesize{$^{\ast}$\ The broad line shape of this feature suggests it may originate
 from a two-phonon process.}
\end{table}
The experiments, reported in Table \ref{Tab2} for both compounds and
in Fig.\ref{fig1} for TiOBr only, show that 10 modes are active in
the $c(aa)c$\ and $c(bb)c$\ in the SP phase (Fig.\ref{fig1}(c)),
and, more importantly, two modes are active in the cross
polarization $c(ab)c$. This is not compatible with the expectation
for P$mm2$. Hence the comparison between the experiments and the
group theoretical analysis clearly shows that of the two low
temperature structures reported in X-ray
crystallography\cite{smaalen2005,sasaki2006}, only the P$2_{1}/m$\
is compatible with the present results.

\begin{figure}[htb]
 \centerline{\includegraphics[width=90mm]{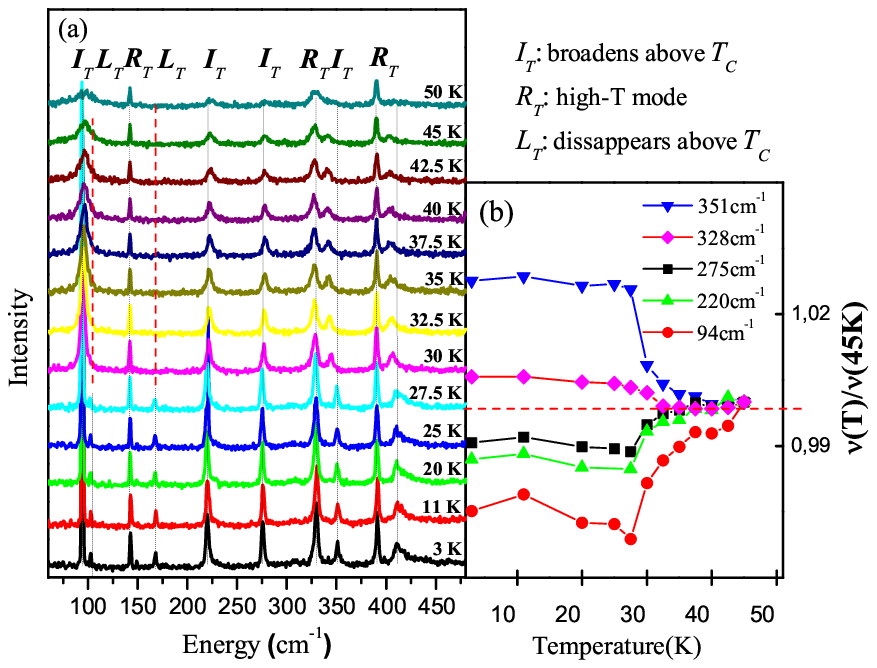}}
 \caption{(Color online) The temperature dependence of the Raman spectrum of TiOBr
is depicted (an offset is added for clarity). The 3 modes present at
all temperatures are denoted by the label $R_{T}$. The modes
characteristic of the low temperature phase (disappearing at
$T_{c1}=28$~K) are labelled $L_{T}$, and the anomalous modes
observed in both the low temperature and the intermediate phase are
labelled $I_{T}$. The right panel (b) shows the behavior of the
frequency of $I_{T}$\ modes, plotted renormalized to their frequency
at 45~K. It is clear that the low-frequency modes shift to higher
energy while the high-frequency modes shift to lower frequency.}
\label{fig3}
\end{figure}

As discussed in the introduction, the presence of three phases in
different temperature intervals for TiOX is now well established
even though the nature of the intermediate phase is still largely
debated\cite{ruck2005,caimi2004,smaalen2005}. The temperature
dependence of the Raman active modes for TiOBr between 3 and 50 K,
is depicted in Fig.\ref{fig3}. In the spin-Peierls phase, as
discussed above, the reduction of the crystal
symmetry\cite{Schon2006} increases the number of Raman active modes.
Increasing the temperature above $T_{c1}$\ a different behavior for
the various low temperature phonons is observed. As shown in
Fig.\ref{fig3}, some of the modes disappear suddenly at $T_{c1}$\
(labeled $L_{T}$), some stay invariant up to the HT phase ($R_{T}$)
and some others undergo a sudden broadening at $T_{c1}$\ and slowly
disappear upon approaching $T_{c2}$\ ($I_{T}$). The polarization
analysis of the Raman modes in the temperature region
$T_{c1}<T<T_{c2}$\ shows that the number of active modes in the
intermediate phase is different from that in both the HT and the SP
phases. The fact that at $T=T_{c1}$\ some of the modes disappear
suddenly while some others do not disappear, strongly suggests that
the crystal symmetry in the intermediate phase is different from
both other phases, and indeed confirms the first-order nature of the
transition at $T_{c1}$.

In the X-ray structure determination \cite{smaalen2005}, the
intermediate incommensurate phase is discussed in two ways. Firstly,
starting from the HT orthorhombic (P$mmn$) and the SP monoclinic
space group (P$2_{1}/m$\ - unique axis in the TiO planes,
$\parallel$\ to $a$), the modulation vector required to explain the
observed incommensurate peaks is two-dimensional for both space
groups. Secondly, starting from another monoclinic space group, with
unique axis perpendicular to the TiO bilayers (P$2/c$), the
modulation vector required is one-dimensional. The latter average
symmetry is considered (in the commensurate variety) in
Fig.\ref{fig2}(a) and Table \ref{Tab3}(a).

In the IP, seven modes are observed in the
$\sigma_{xx},\sigma_{yy}$\ and $\sigma_{zz}$\ geometry on both
compounds (see Table \ref{Tab2}(b)), and none in the $\sigma_{xy}$\
geometry. This appears to be compatible with all the space groups
considered, and also with the monoclinic group with unique axis
perpendicular to the TiO planes (Table \ref{Tab3}(a)). Even though
from the evidence it is not possible to rule out any of the other
symmetries discussed, the conjecture that in the intermediate
incommensurate phase the average crystal symmetry is already
reduced, supports the description of the intermediate phase as a
monoclinic group with a one-dimensional
modulation\cite{smaalen2005}, and moreover it explains the anomalous
first-order character of the spin-Peierls transition at $T_{c1}$.
\begin{figure}[htb]
 \centerline{\includegraphics[width=80mm]{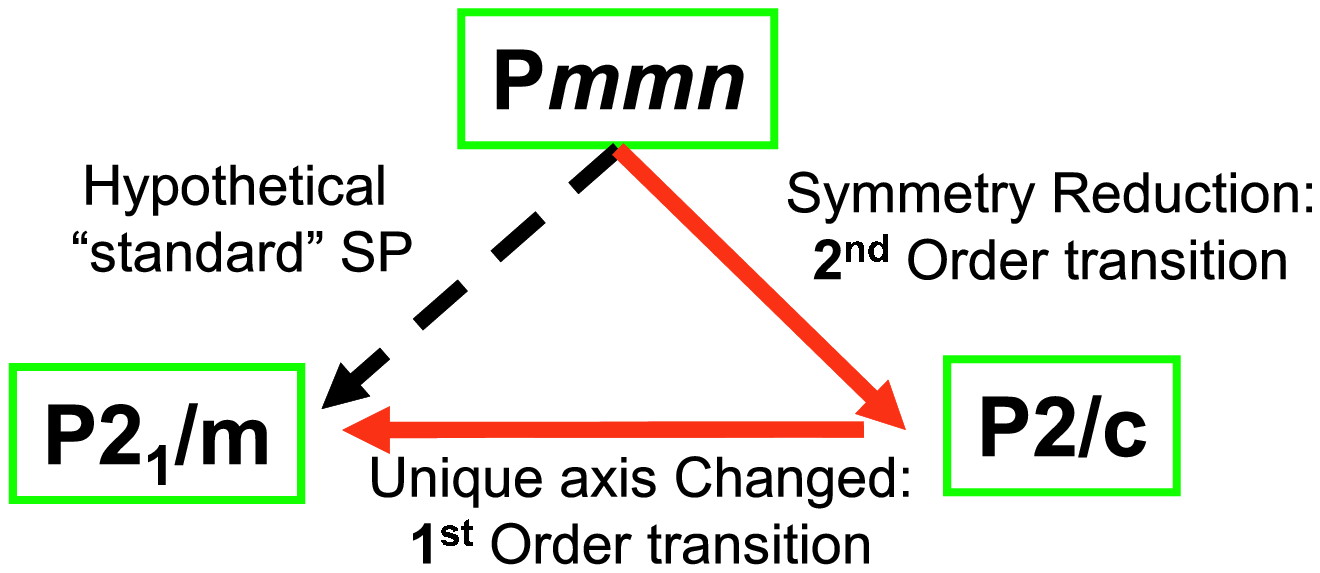}}
 \caption{(Color online) The average crystal symmetry of the intermediate phase
 is proposed to be monoclinic with the unique axis parallel to the $c-$axis of the orthorhombic phase.
 Hence the low temperature space group is not a subgroup of the intermediate phase,
 and the transition to the spin-Peierls phase is consequently of first order.}
 \label{fig4}
\end{figure}
The diagram shown in Fig.\ref{fig4} aims to visualize that the space
group in the spin-Peierls state (P$2_{1}/m$) is a subgroup of the
high temperature P$mmn$\ group, but not a subgroup of any of the
possible intermediate phase space groups suggested (possible
P$2/c$). This requires the phase transition at $T_{c1}$ to be of
first order, instead of having the conventional spin-Peierls
second-order character.

Let us return to Fig.\ref{fig3}(b) to discuss another intriguing
vibrational feature of the intermediate phase. Among the modes
characterizing the intermediate phase ($I_{T}$), the ones at low
frequency shift to higher energy approaching $T_{c2}$, while the
ones at high frequency move to lower energy, seemingly converging to
a central frequency ($\simeq$300~\cm\ for both TiOCl and TiOBr).
This seems to indicate an interaction of the phonons with some
excitation around 300~\cm. Most likely this is in fact arising from
a strong, thermally activated coupling of the lattice with the
magnetic excitations, and is consistent with the pseudo-spin gap
observed in NMR experiments\cite{imai2003,bak2007} of $\approx$430 K
($\simeq$300 \cm).

\subsection{Magnetic Interactions}
As discussed in the introduction, due to the shape of the singly
occupied 3$d$\ orbital, the main magnetic exchange interaction
between the spins on the Ti ions is along the crystallographic
$b-$direction.
\begin{figure}[htb]
 \centerline{\includegraphics[width=90mm]{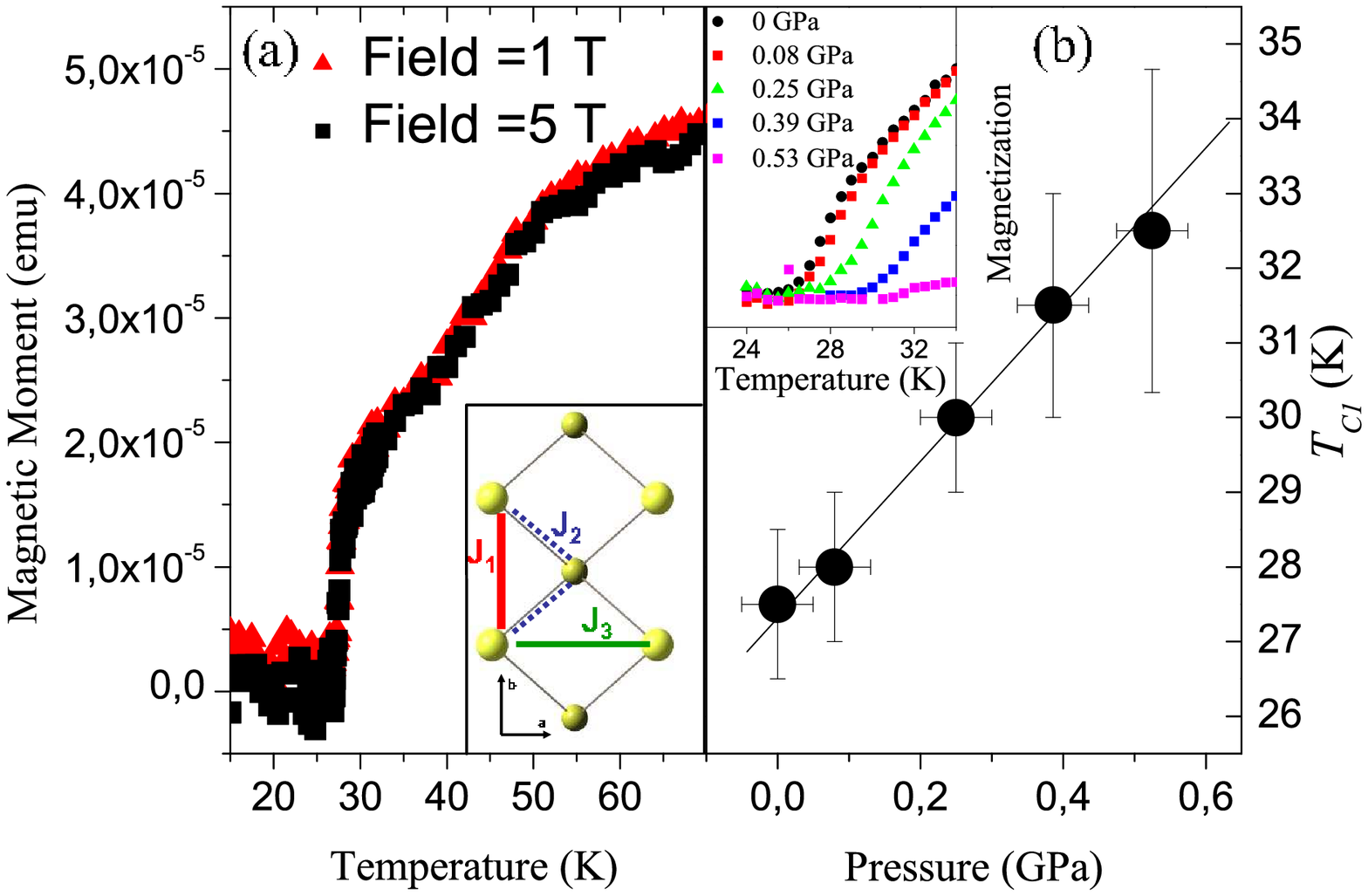}}
 \caption{(Color online) (a) Magnetization as a function of temperature
measured with fields 1 T and 5 T (the magnetization measured at 1 T
is multiplied by a factor of 5 to evidence the linearity). The inset
shows the main magnetic interactions (see text). (b) Pressure
dependence of $T_{c1}$. The transition temperature for transition to
the spin-Peierls phase increases with increasing pressure. The inset
shows the magnetization versus the temperature after subtracting the
background signal coming from the pressure cell.}
 \label{fig5}
\end{figure}
This, however, is not the only effective magnetic interaction. In
fact, one also expects a superexchange interaction between nearest
and next-nearest neighbor chains ($J_{2}$\ and $J_{3}$\ in the
insert of Fig.\ref{fig5}(a))\cite{Roberto}. The situation of TiOX is
made more interesting by the frustrated geometry of the interchain
interaction, where the magnetic coupling $J_{2}$\ between adjacent
chains is frustrated and the exchange energies can not be
simultaneously minimized. Table V reports the exchange interaction
values for the three possible magnetic interactions calculated for
TiOBr. These magnetic interactions were computed with a DFT Broken
symmetry approach\cite{Nood79} using an atom cluster including the
two interacting atoms and all the surrounding ligand atoms, in
addition the first shell of Ti$^{3+}$\ ions was replaced by
Al$^{3+}$\ ions and also included in the cluster. The calculations
were performed with the Gaussian03 package\cite{Frish04} using the
hybrid exchange-correlation functional B3LYP\cite{Beck93} and the
6-3111G* basisset.

\begin{table}[htb]
 \caption{Calculated Exchange interactions in TiOBr}
 \label{Tab5}
 \begin{ruledtabular}
 \begin{tabular}{c}
TiOBr\\
$J_{1}=-250$~K \\
$J_{2}=-46.99$~K\\
$J_{3}=11.96$~K\\
\end{tabular}
 \end{ruledtabular}
 \end{table}

Although the computed value for the magnetic interaction along the
$b-$axis is half of the value obtained from the magnetic
susceptibility fitted with a Bonner-Fisher curve accounting for a
one-dimensional Heisenberg chain, it is possible to extract some
conclusions from the ab-initio computations. The most interesting
outcome of the results is that in addition to the magnetic
interaction along the $b-$axis, there is a relevant interchain
interaction ($J_{1}/J_{2}= 5.3$) in TiOBr. Firstly, this explains
the substantial deviation of the Bonner-Fisher fit from the magnetic
susceptibility even at temperature higher than $T_{c2}$. Secondly,
the presence of an interchain interaction, together with the
inherent frustrated geometry of the bilayer structure, was already
proposed in literature\cite{ruck2005} in order to explain the
intermediate phase and its structural incommensurability.

The two competing exchange interactions $J_{1}$\ and $J_{2}$\ have
different origins: the first arises from direct exchange between Ti
ions, while the second is mostly due to the superexchange
interaction through the oxygen ions\cite{Roberto}. Thus, the two
exchange constants are expected to depend differently on the
structural changes induced by hydrostatic pressure, $J_{1}$\ should
increase with hydrostatic pressure (increases strongly with
decreasing the distance between the Ti ions), while $J_{2}$\ is
presumably weakly affected due only to small changes in the
Ti--O--Ti angle (the compressibility estimated from the lattice
dynamics simulation is similar along the $a$\ and $b$\
crystallographic directions). The stability of the fully dimerized
state is reduced by the presence of an interchain coupling, so that
$T_{c1}$\ is expected to be correlated to $J_{1}/J_{2}$. Pressure
dependent magnetic experiments have been performed to monitor the
change of $T_{c1}$\ upon increasing hydrostatic pressure. The main
results, shown in Fig.\ref{fig5}, indeed is consistent with this
expectation: $T_{c1}$\ increases linearly with pressure;
unfortunately it is not possible to address the behavior of
$T_{c2}$\ from the present measurements.

\subsection{Electronic Excitations and Comparison with VOCl}
The nature of the complex phase diagram of TiOX was originally
tentatively ascribed to the interplay of spin, lattice and orbital
degrees of freedom\cite{caimi2004}. Only recently, infrared
spectroscopy supported by cluster calculations excluded a ground
state degeneracy of the Ti $d$\ orbitals for TiOCl, hence suggesting
that orbital fluctuations can not play an important role in the
formation of the anomalous incommensurate
phase\cite{ruck2005long,Zach2006}. Since the agreement between the
previous cluster calculations and the experimental results is not
quantitative, the energy of the lowest $3d$\ excited level is not
accurately known, not allowing to discard the possibility of an
almost degenerate ground state. For this reason a more formal
cluster calculation has been performed using an embedded cluster
approach. In this approach a TiO2Cl4 cluster was treated explicitly
with a CASSCF/CASPT2 quantum chemistry calculation. This cluster was
surrounded by eight Ti$^{3+}$\ TIP potentials in order to account
for the electrostatic interaction of the cluster atoms with the
shell of the first neighboring atoms. Finally, the cluster is
embedded in a distribution of punctual charges fitting the
Madelung's potential produced by the rest of the crystal inside the
cluster region. The calculations were performed using the MOLCAS
quantum chemistry package\cite{Karl2003} with a triple quality basis
set; for the Ti atom polarization functions were also included.
\begin{table}[htb]
 \caption{Crystal field splitting of 3$d^{1}$\ Ti$^{3+}$\ in TiOCl and TiOBr (eV).}
 \label{Tab4}
 \begin{ruledtabular}
 \begin{tabular}{lll}
& TiOCl & TiOBr\\
\hline
$xy$& 0.29-0.29 & 0.29-0.30\\
$xz$& 0.66-0.68 & 0.65-0.67\\
$yz$& 1.59-1.68 & 1.48-1.43\\
$x^{2}-r^{2}$& 2.30-2.37& 2.21-2.29\\
\end{tabular}
 \end{ruledtabular}
 \end{table}
The calculations reported in Table \ref{Tab4},
confirmed the previously reported result\cite{ruck2005long} for
both TiOCl and TiOBr. The first excited state $d_{xy}$\ is at
0.29-0.3 eV ($>3000$~K) for both compounds, therefore the orbital
degrees of freedom are completely quenched at temperatures close
to the phase transition.

A comparison with the isostructural compound VOCl has been carried
out to confirm that the phase transitions of the TiOX compounds are
intimately related to the unpaired S=1/2 spin of the Ti ions. The
V$^{3+}$\ ions have a 3$d^{2}$\ electronic configuration. Each ion
carries two unpaired electrons in the external d shell, and has a
total spin of 1. The crystal field environment of V$^{3+}$\ ions in
VOCl is similar to that of Ti$^{3+}$\ in TiOX, suggesting that the
splitting of the degenerate d orbital could be comparable. The
electrons occupy the two lowest $t_{2g}$\ orbitals, of
$d_{y^2-z^2}$\ (responsible for the main exchange interaction in
TiOX) and $d_{xy}$\ symmetry respectively. Where the lobes of the
latter point roughly towards the Ti$^{3+}$\ ions of the nearest
chain (Table \ref{Tab4}). It is therefore reasonable to expect that
the occupation of the $d_{xy}$\ orbital in VOCl leads to a
substantial direct exchange interaction between ions in different
chains in VOCl and thus favors a two-dimensional antiferromagnetic
order. Indeed, the magnetic susceptibility is isotropic at high
temperatures and well described by a quadratic two-dimensional
Heisenberg model, and at $T_{N}=80$~K VOCl undergoes a phase
transition to a two-dimensional antiferromagnet\cite{wied84}.

\begin{figure}[htb]
 \centerline{\includegraphics[width=60mm]{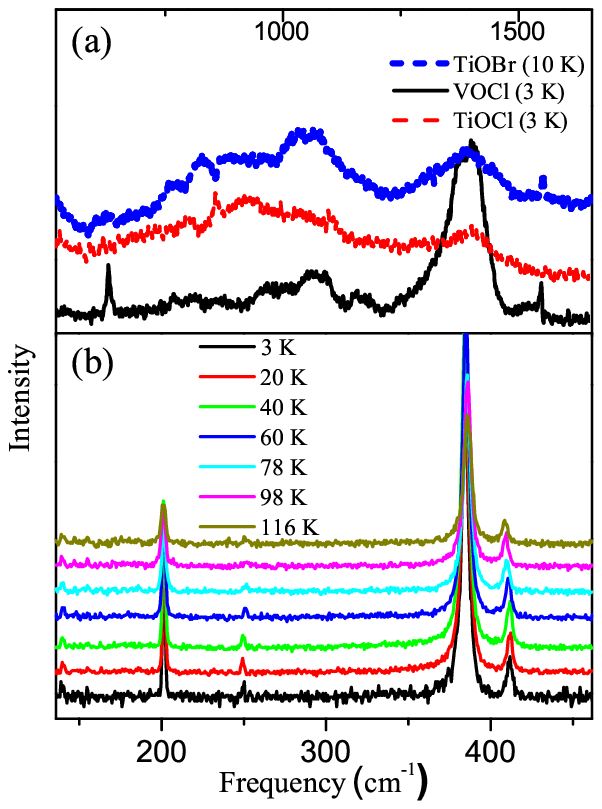}}
 \caption{(Color online) Raman scattering features of VOCl.
(a) High energy scattering of TiOCl/Br and VOCl, and (b)
temperature dependence of the vibrational scattering features of
VOCl. No symmetry changes are observed at $T_{N}=80$~K.}
 \label{fig6}
\end{figure}

The space group of VOCl at room temperature is the same as that of
TiOX in the high temperature phase (P$mmn$), and, as discussed in
the previous section, three $A_{g}$\ modes are expected to be Raman
active. As shown in Fig.\ref{fig6}(b), three phonons are observed
throughout the full temperature range ($3-300$~K), and no changes are observed at
$T_{N}$. The modes observed are consistent with the prediction of
lattice dynamics calculations (Table \ref{Tab1}).

In the energy region from 600 to 1500~\cm, both TiOBr and TiOCl show
a similar highly structured broad scattering continuum, as already
reported in literature\cite{Lemmens2003,lemmens2005}. The fact that
the energy range of the anomalous feature is consistent with the
magnetic exchange constant in TiOCl (J=660~K) suggested at first an
interpretation in terms of two-magnon Raman scattering
\cite{Lemmens2003}. Later it was shown that the exchange constant
estimated for TiOBr is considerably smaller (J=406~K) with respect
to that of TiOCl while the high energy scattering stays roughly at
the same frequency. Even though the authors of
ref.\cite{lemmens2005} still assigned the scattering continuum to
magnon processes, it seems clear taht the considerably smaller
exchange interaction in the Br compound (J=406~K) falsifies this
interpretation and that magnon scattering is not at the origin of
the high energy scattering of the two compounds. Furthermore, the
cluster calculation (Table \ref{Tab4}) clearly shows that no excited
crystal field state is present in the energy interval considered,
ruling out a possible orbital origin for the continuum. These
observations are further strengthened by the observation of a
similar continuum scattering in VOCl (see fig. \ref{fig6}(a)) which
has a different magnetic and electronic nature. Therefore, the high
energy scattering has most likely a vibrational origin. The lattice
dynamics calculations, confirmed by the experiments, show that a
"high" energy mode ($\simeq$600~\cm) of symmetry $B_{3g}$\ (Table
\ref{Tab1}) is expected to be Raman active in the $\sigma_{yz}$\
polarization. Looking back at Fig.\ref{fig1}, the inset shows the
measurements performed with the optical axis parallel to the TiOX
plane, where the expected mode is observed at 598~\cm. The two
phonon process related to this last intense mode is in the energy
range of the anomalous scattering feature and has symmetry $A_{g}$\
($B_{3g} \otimes B_{3g}$). The nature of the anomalies observed is
therefore tentatively ascribed to a multiple-phonon process. Further
detailed investigations of lattice dynamics are needed to clarify
this issue.

\section{Conclusion}

The symmetry of the different phases has been discussed on the basis
of inelastic light scattering experiments. The high temperature
Raman experiments are in good agreement with the prediction of the
group theoretical analysis (apart from one broad mode which is
ascribed to pre-transitional fluctuations). Comparing group
theoretical analysis with the polarized Raman spectra clarifies the
symmetry of the spin-Peierls phase and shows that the average
symmetry of the incommensurate phase is different from both the high
temperature and the SP phases. The conjecture that the intermediate
phase is compatible with a different monoclinic symmetry (unique
axis perpendicular to the TiO planes) could explain the anomalous
first-order character of the transition to the spin-Peierls phase.
Moreover, an anomalous behavior of the phonons characterizing the
intermediate phase is interpreted as evidencing an important
spin-lattice coupling. The susceptibility measurements of TiOBr show
that $T_{c1}$\ increases with pressure, which is ascribed to the
different pressure dependence of intrachain and interchain
interactions. Finally, we compared the TiOX compounds with the
"isostructural" VOCl. The presence of the same anomalous high energy
scattering feature in all the compounds suggests that this feature
has a vibrational origin rather than a magnetic or electronic one.

{\em Acknowledgements}\ \ \
The authors are grateful to Maxim Mostovoy, Michiel van der Vegte,
Paul de Boeij, Daniel Khomskii, Iberio Moreira and Markus
Gr\"uninger for valuable and insightful discussions. This work was
partially supported by the Stichting voor Fundamenteel Onderzoek
der Materie [FOM, financially supported by the Nederlandse
Organisatie voor Wetenschappelijk Onderzoek (NWO)], and by the
German Science Foundation (DFG).

\section{Appendix: Details of the spring model calculation}
The spring model calculation reported in the paper, was carried out
using the software for lattice-dynamical calculation
UNISOFT\cite{esc92} (release 3.05). In the calculations the Born-von
Karman model was used; here the force constants are treated as model
parameters and they are not interpreted in terms of a special
interatomic potential. Only short range interactions between nearest
neighbor ions are taken into account. Considering the forces to be
central forces, the number of parameters is reduced to two for each
atomic interaction: the longitudinal and transversal forces
respectively defined as $L=\frac{d^2V(\bar{r}_{i,j})}{dr^2}$ and
$T=\frac{1}{r}\frac{dV(\bar{r}_{i,j})}{dr}$. A custom made program
was interfaced with UNISOFT to optimize the elastic constants. Our
program proceeded scanning the $n$ dimensional space ($n$ = number
of parameters) with a discrete grid, to minimize the squared
difference between the calculated phonon frequencies and the
measured experimental frequencies for TiOBr, taken from both Raman
and infrared spectroscopy. The phonon frequencies of TiOCl and VOCl
were obtained using the elastic constants optimized for TiOBr and
substituting the appropriate ionic masses. The optimized force
constants between different atoms are reported in $N/m$ in the
following Table.

\begin{table}[htb]
\caption{Elastic constants used in the spring model calculation. The
label numbers refer to Fig. \ref{add1}, while the letters refer to
the different inequivalent positions of the ions in the crystal.}
 \begin{ruledtabular}
\begin{tabular}{cccc}
Number & Ions  & Longitudinal (L) ($N/m$) & Transversal (T) ($N/m$)\\
1 & Ti(a)-Ti(b)& 18.5 & 32.7\\
2 & Ti(a)-O(a) & 18.5 & 11.1\\
3 & Ti(a)-O(b) & 53.1 & 9.5\\
4 & Ti(a)-X(a)& 29.0 & 4.4\\
5 & O(a)-O(b)  & 20.6 & 7.3\\
6 & X(a)-O(a) & 18.5 & 3.5\\
7 & X(a)-X(b)& 11.7 & 0.7\\
 \end{tabular}
 \end{ruledtabular}
\label{Tadd1}
\end{table}

%\begin{figure}[htb]
% \includegraphics[width=90mm]{appendix.eps}
% \caption{(Color online) Sketch of the bilayer structure (b) and of the interactions introduced in the spring model calculation (a).}
% \label{add1}
%\end{figure}

\bibliography{bibliografia1}

\begin{thebibliography}{31}
\expandafter\ifx\csname natexlab\endcsname\relax\def\natexlab#1{#1}\fi
\expandafter\ifx\csname bibnamefont\endcsname\relax
  \def\bibnamefont#1{#1}\fi
\expandafter\ifx\csname bibfnamefont\endcsname\relax
  \def\bibfnamefont#1{#1}\fi
\expandafter\ifx\csname citenamefont\endcsname\relax
  \def\citenamefont#1{#1}\fi
\expandafter\ifx\csname url\endcsname\relax
  \def\url#1{\texttt{#1}}\fi
\expandafter\ifx\csname urlprefix\endcsname\relax\def\urlprefix{URL }\fi
\providecommand{\bibinfo}[2]{#2}
\providecommand{\eprint}[2][]{\url{#2}}

\bibitem[{\citenamefont{Imada et~al.}(1998)\citenamefont{Imada, Fujimori, and
  Tokura}}]{Imad98}
\bibinfo{author}{\bibfnamefont{M.}~\bibnamefont{Imada}},
  \bibinfo{author}{\bibfnamefont{A.}~\bibnamefont{Fujimori}}, \bibnamefont{and}
  \bibinfo{author}{\bibfnamefont{Y.}~\bibnamefont{Tokura}},
  \bibinfo{journal}{Rev. Mod. Phys.} \textbf{\bibinfo{volume}{70}},
  \bibinfo{pages}{1039} (\bibinfo{year}{1998}).

\bibitem[{\citenamefont{Dagotto}(1999)}]{Dag99}
\bibinfo{author}{\bibfnamefont{E.}~\bibnamefont{Dagotto}},
  \bibinfo{journal}{Rep. Prog. Phys.} \textbf{\bibinfo{volume}{62}},
  \bibinfo{pages}{1525} (\bibinfo{year}{1999}).

\bibitem[{\citenamefont{Kataev et~al.}(2003)\citenamefont{Kataev, Baier,
  M$\ddot{\textrm{o}}$ller, Jongen, Meyer, , and Freimuth}}]{kataev2003}
\bibinfo{author}{\bibfnamefont{V.}~\bibnamefont{Kataev}},
  \bibinfo{author}{\bibfnamefont{J.}~\bibnamefont{Baier}},
  \bibinfo{author}{\bibfnamefont{A.}~\bibnamefont{M$\ddot{\textrm{o}}$ller}},
  \bibinfo{author}{\bibfnamefont{L.}~\bibnamefont{Jongen}},
  \bibinfo{author}{\bibfnamefont{G.}~\bibnamefont{Meyer}}, , \bibnamefont{and}
  \bibinfo{author}{\bibfnamefont{A.}~\bibnamefont{Freimuth}},
  \bibinfo{journal}{Phys. Rev. B} \textbf{\bibinfo{volume}{68}},
  \bibinfo{pages}{140405} (\bibinfo{year}{2003}).

\bibitem[{\citenamefont{Imai and Choub}(2003)}]{imai2003}
\bibinfo{author}{\bibfnamefont{T.}~\bibnamefont{Imai}} \bibnamefont{and}
  \bibinfo{author}{\bibfnamefont{F.~C.} \bibnamefont{Choub}},
  \bibinfo{journal}{cond-mat} \textbf{\bibinfo{volume}{0301425}}
  (\bibinfo{year}{2003}),
  \urlprefix\url{http://xxx.lanl.gov/abs/cond-mat/0301425}.

\bibitem[{\citenamefont{Maule et~al.}(1988)\citenamefont{Maule, Tothill,
  Strange, and Wilson}}]{Maule88}
\bibinfo{author}{\bibfnamefont{C.~H.} \bibnamefont{Maule}},
  \bibinfo{author}{\bibfnamefont{J.~N.} \bibnamefont{Tothill}},
  \bibinfo{author}{\bibfnamefont{P.}~\bibnamefont{Strange}}, \bibnamefont{and}
  \bibinfo{author}{\bibfnamefont{J.~A.} \bibnamefont{Wilson}},
  \bibinfo{journal}{J. Phys. C} \textbf{\bibinfo{volume}{21}},
  \bibinfo{pages}{2153} (\bibinfo{year}{1988}).

\bibitem[{\citenamefont{Seidel et~al.}(2003)\citenamefont{Seidel, Marianetti,
  Chou, Ceder, and Lee}}]{seidel2003}
\bibinfo{author}{\bibfnamefont{A.}~\bibnamefont{Seidel}},
  \bibinfo{author}{\bibfnamefont{C.~A.} \bibnamefont{Marianetti}},
  \bibinfo{author}{\bibfnamefont{F.~C.} \bibnamefont{Chou}},
  \bibinfo{author}{\bibfnamefont{G.}~\bibnamefont{Ceder}}, \bibnamefont{and}
  \bibinfo{author}{\bibfnamefont{P.~A.} \bibnamefont{Lee}},
  \bibinfo{journal}{Phys. Rev. B} \textbf{\bibinfo{volume}{67}},
  \bibinfo{pages}{020405} (\bibinfo{year}{2003}).

\bibitem[{\citenamefont{Caimi et~al.}(2004{\natexlab{a}})\citenamefont{Caimi,
  Degiorgi, Lemmens, and Chou}}]{caimi2004}
\bibinfo{author}{\bibfnamefont{G.}~\bibnamefont{Caimi}},
  \bibinfo{author}{\bibfnamefont{L.}~\bibnamefont{Degiorgi}},
  \bibinfo{author}{\bibfnamefont{P.}~\bibnamefont{Lemmens}}, \bibnamefont{and}
  \bibinfo{author}{\bibfnamefont{F.~C.} \bibnamefont{Chou}},
  \bibinfo{journal}{J. Phys. Cond. Mat.} \textbf{\bibinfo{volume}{16}},
  \bibinfo{pages}{5583} (\bibinfo{year}{2004}{\natexlab{a}}).

\bibitem[{\citenamefont{Beynon and Wilson}(1993)}]{Beynon1993}
\bibinfo{author}{\bibfnamefont{R.~J.} \bibnamefont{Beynon}} \bibnamefont{and}
  \bibinfo{author}{\bibfnamefont{J.~A.} \bibnamefont{Wilson}},
  \bibinfo{journal}{J. Phys. Cond. Mat.} \textbf{\bibinfo{volume}{5}},
  \bibinfo{pages}{1983} (\bibinfo{year}{1993}).

\bibitem[{\citenamefont{Caimi et~al.}(2004{\natexlab{b}})\citenamefont{Caimi,
  Degiorgi, Kovaleva, Lemmens, and Chou}}]{caimi2004-1}
\bibinfo{author}{\bibfnamefont{G.}~\bibnamefont{Caimi}},
  \bibinfo{author}{\bibfnamefont{L.}~\bibnamefont{Degiorgi}},
  \bibinfo{author}{\bibfnamefont{N.~N.} \bibnamefont{Kovaleva}},
  \bibinfo{author}{\bibfnamefont{P.}~\bibnamefont{Lemmens}}, \bibnamefont{and}
  \bibinfo{author}{\bibfnamefont{F.~C.} \bibnamefont{Chou}},
  \bibinfo{journal}{Phys. Rev. B} \textbf{\bibinfo{volume}{69}},
  \bibinfo{pages}{125108} (\bibinfo{year}{2004}{\natexlab{b}}).

\bibitem[{\citenamefont{Shaz et~al.}(2005)\citenamefont{Shaz, van Smaalen,
  Palatinus, Hoinkis, Klemm, Horn, and Claessen}}]{shaz}
\bibinfo{author}{\bibfnamefont{M.}~\bibnamefont{Shaz}},
  \bibinfo{author}{\bibfnamefont{S.}~\bibnamefont{van Smaalen}},
  \bibinfo{author}{\bibfnamefont{L.}~\bibnamefont{Palatinus}},
  \bibinfo{author}{\bibfnamefont{M.}~\bibnamefont{Hoinkis}},
  \bibinfo{author}{\bibfnamefont{M.}~\bibnamefont{Klemm}},
  \bibinfo{author}{\bibfnamefont{S.}~\bibnamefont{Horn}}, \bibnamefont{and}
  \bibinfo{author}{\bibfnamefont{R.}~\bibnamefont{Claessen}},
  \bibinfo{journal}{Phys. Rev. B} \textbf{\bibinfo{volume}{71}},
  \bibinfo{pages}{100405} (\bibinfo{year}{2005}).

\bibitem[{\citenamefont{Hemberger et~al.}(2005)\citenamefont{Hemberger,
  Hoinkis, Klemm, Sing, Claessen, Horn, and Loidl}}]{hemberger}
\bibinfo{author}{\bibfnamefont{J.}~\bibnamefont{Hemberger}},
  \bibinfo{author}{\bibfnamefont{M.}~\bibnamefont{Hoinkis}},
  \bibinfo{author}{\bibfnamefont{M.}~\bibnamefont{Klemm}},
  \bibinfo{author}{\bibfnamefont{M.}~\bibnamefont{Sing}},
  \bibinfo{author}{\bibfnamefont{R.}~\bibnamefont{Claessen}},
  \bibinfo{author}{\bibfnamefont{S.}~\bibnamefont{Horn}}, \bibnamefont{and}
  \bibinfo{author}{\bibfnamefont{A.}~\bibnamefont{Loidl}},
  \bibinfo{journal}{Phys. Rev. B} \textbf{\bibinfo{volume}{72}},
  \bibinfo{pages}{012420} (\bibinfo{year}{2005}).

\bibitem[{\citenamefont{R$\ddot{\textrm{u}}$ckamp
  et~al.}(2005{\natexlab{a}})\citenamefont{R$\ddot{\textrm{u}}$ckamp, Baier,
  Kriener, Haverkort, Lorenz, Uhrig, Jongen, M$\ddot{\textrm{o}}$ller, Meyer,
  and Gr$\ddot{\textrm{u}}$ninger}}]{ruck2005}
\bibinfo{author}{\bibfnamefont{R.}~\bibnamefont{R$\ddot{\textrm{u}}$ckamp}},
  \bibinfo{author}{\bibfnamefont{J.}~\bibnamefont{Baier}},
  \bibinfo{author}{\bibfnamefont{M.}~\bibnamefont{Kriener}},
  \bibinfo{author}{\bibfnamefont{M.~W.} \bibnamefont{Haverkort}},
  \bibinfo{author}{\bibfnamefont{T.}~\bibnamefont{Lorenz}},
  \bibinfo{author}{\bibfnamefont{G.~S.} \bibnamefont{Uhrig}},
  \bibinfo{author}{\bibfnamefont{L.}~\bibnamefont{Jongen}},
  \bibinfo{author}{\bibfnamefont{A.}~\bibnamefont{M$\ddot{\textrm{o}}$ller}},
  \bibinfo{author}{\bibfnamefont{G.}~\bibnamefont{Meyer}}, \bibnamefont{and}
  \bibinfo{author}{\bibfnamefont{M.}~\bibnamefont{Gr$\ddot{\textrm{u}}$ninger}%
}, \bibinfo{journal}{Phys. Rev. Lett.} \textbf{\bibinfo{volume}{95}},
  \bibinfo{pages}{097203} (\bibinfo{year}{2005}{\natexlab{a}}).

\bibitem[{\citenamefont{Lemmens et~al.}(2004)\citenamefont{Lemmens, Choi,
  Caimi, Degiorgi, Kovaleva, Seidel, and Chou}}]{Lemmens2003}
\bibinfo{author}{\bibfnamefont{P.}~\bibnamefont{Lemmens}},
  \bibinfo{author}{\bibfnamefont{K.~Y.} \bibnamefont{Choi}},
  \bibinfo{author}{\bibfnamefont{G.}~\bibnamefont{Caimi}},
  \bibinfo{author}{\bibfnamefont{L.}~\bibnamefont{Degiorgi}},
  \bibinfo{author}{\bibfnamefont{N.~N.} \bibnamefont{Kovaleva}},
  \bibinfo{author}{\bibfnamefont{A.}~\bibnamefont{Seidel}}, \bibnamefont{and}
  \bibinfo{author}{\bibfnamefont{F.~C.} \bibnamefont{Chou}},
  \bibinfo{journal}{Phys. Rev. B} \textbf{\bibinfo{volume}{70}},
  \bibinfo{pages}{134429} (\bibinfo{year}{2004}).

\bibitem[{\citenamefont{Palatinus et~al.}(2005)\citenamefont{Palatinus,
  Schoenleber, and van Smaalen}}]{palatinus2005}
\bibinfo{author}{\bibfnamefont{L.}~\bibnamefont{Palatinus}},
  \bibinfo{author}{\bibfnamefont{A.}~\bibnamefont{Schoenleber}},
  \bibnamefont{and} \bibinfo{author}{\bibfnamefont{S.}~\bibnamefont{van
  Smaalen}}, \bibinfo{journal}{Acta Crystallogr. Sect. C}
  \textbf{\bibinfo{volume}{61}}, \bibinfo{pages}{148} (\bibinfo{year}{2005}).

\bibitem[{\citenamefont{van Smaalen et~al.}(2005)\citenamefont{van Smaalen,
  Palatinus, and Schoenleber}}]{smaalen2005}
\bibinfo{author}{\bibfnamefont{S.}~\bibnamefont{van Smaalen}},
  \bibinfo{author}{\bibfnamefont{L.}~\bibnamefont{Palatinus}},
  \bibnamefont{and}
  \bibinfo{author}{\bibfnamefont{A.}~\bibnamefont{Schoenleber}},
  \bibinfo{journal}{Phys. Rev. B} \textbf{\bibinfo{volume}{72}},
  \bibinfo{pages}{020105(R)} (\bibinfo{year}{2005}).

\bibitem[{\citenamefont{Schoenleber et~al.}(2006)\citenamefont{Schoenleber, van
  Smaalen, and Palatinus}}]{Schon2006}
\bibinfo{author}{\bibfnamefont{A.}~\bibnamefont{Schoenleber}},
  \bibinfo{author}{\bibfnamefont{S.}~\bibnamefont{van Smaalen}},
  \bibnamefont{and}
  \bibinfo{author}{\bibfnamefont{L.}~\bibnamefont{Palatinus}},
  \bibinfo{journal}{Phys. Rev. B} \textbf{\bibinfo{volume}{73}},
  \bibinfo{pages}{214410} (\bibinfo{year}{2006}).

\bibitem[{\citenamefont{Krimmel et~al.}(2006)\citenamefont{Krimmel, Strempfer,
  Bohnenbuck, Keimer, Hoinkis, Klemm, Horn, Loidl, Sing, Claessen
  et~al.}}]{krim2006}
\bibinfo{author}{\bibfnamefont{A.}~\bibnamefont{Krimmel}},
  \bibinfo{author}{\bibfnamefont{J.}~\bibnamefont{Strempfer}},
  \bibinfo{author}{\bibfnamefont{B.}~\bibnamefont{Bohnenbuck}},
  \bibinfo{author}{\bibfnamefont{B.}~\bibnamefont{Keimer}},
  \bibinfo{author}{\bibfnamefont{M.}~\bibnamefont{Hoinkis}},
  \bibinfo{author}{\bibfnamefont{M.}~\bibnamefont{Klemm}},
  \bibinfo{author}{\bibfnamefont{S.}~\bibnamefont{Horn}},
  \bibinfo{author}{\bibfnamefont{A.}~\bibnamefont{Loidl}},
  \bibinfo{author}{\bibfnamefont{M.}~\bibnamefont{Sing}},
  \bibinfo{author}{\bibfnamefont{R.}~\bibnamefont{Claessen}},
  \bibnamefont{et~al.}, \bibinfo{journal}{Phys. Rev. B}
  \textbf{\bibinfo{volume}{73}}, \bibinfo{pages}{172413}
  (\bibinfo{year}{2006}).

\bibitem[{\citenamefont{Kuntscher et~al.}(2006)\citenamefont{Kuntscher, Frank,
  Pashkin, Hoinkis, Klemm, Sing, Horn, and Claessen}}]{kun2006}
\bibinfo{author}{\bibfnamefont{C.~A.} \bibnamefont{Kuntscher}},
  \bibinfo{author}{\bibfnamefont{S.}~\bibnamefont{Frank}},
  \bibinfo{author}{\bibfnamefont{A.}~\bibnamefont{Pashkin}},
  \bibinfo{author}{\bibfnamefont{M.}~\bibnamefont{Hoinkis}},
  \bibinfo{author}{\bibfnamefont{M.}~\bibnamefont{Klemm}},
  \bibinfo{author}{\bibfnamefont{M.}~\bibnamefont{Sing}},
  \bibinfo{author}{\bibfnamefont{S.}~\bibnamefont{Horn}}, \bibnamefont{and}
  \bibinfo{author}{\bibfnamefont{R.}~\bibnamefont{Claessen}},
  \bibinfo{journal}{Phys. Rev. B} \textbf{\bibinfo{volume}{74}},
  \bibinfo{pages}{184402} (\bibinfo{year}{2006}).

\bibitem[{\citenamefont{Lemmens et~al.}(2005)\citenamefont{Lemmens, Choi,
  Valenti, Saha-Dasgupta, Abel, Lee, and Chou}}]{lemmens2005}
\bibinfo{author}{\bibfnamefont{P.}~\bibnamefont{Lemmens}},
  \bibinfo{author}{\bibfnamefont{K.~Y.} \bibnamefont{Choi}},
  \bibinfo{author}{\bibfnamefont{R.}~\bibnamefont{Valenti}},
  \bibinfo{author}{\bibfnamefont{T.}~\bibnamefont{Saha-Dasgupta}},
  \bibinfo{author}{\bibfnamefont{E.}~\bibnamefont{Abel}},
  \bibinfo{author}{\bibfnamefont{Y.~S.} \bibnamefont{Lee}}, \bibnamefont{and}
  \bibinfo{author}{\bibfnamefont{F.~C.} \bibnamefont{Chou}},
  \bibinfo{journal}{New Journal of Pysics} \textbf{\bibinfo{volume}{7}},
  \bibinfo{pages}{74} (\bibinfo{year}{2005}).

\bibitem[{\citenamefont{Rousseau et~al.}(1981)\citenamefont{Rousseau, Bauman,
  and Porto}}]{Rou1981}
\bibinfo{author}{\bibfnamefont{D.~L.} \bibnamefont{Rousseau}},
  \bibinfo{author}{\bibfnamefont{R.~P.} \bibnamefont{Bauman}},
  \bibnamefont{and} \bibinfo{author}{\bibfnamefont{S.~P.~S.}
  \bibnamefont{Porto}}, \bibinfo{journal}{Journal of Raman Spectroscopy}
  \textbf{\bibinfo{volume}{10}}, \bibinfo{pages}{253} (\bibinfo{year}{1981}).

\bibitem[{\citenamefont{Sasaki et~al.}(2006)\citenamefont{Sasaki, Nagai, Kato,
  Mizumaki, Asaka, Takata, Matsui, Sawa, and Akimitsu}}]{sasaki2006}
\bibinfo{author}{\bibfnamefont{T.}~\bibnamefont{Sasaki}},
  \bibinfo{author}{\bibfnamefont{T.}~\bibnamefont{Nagai}},
  \bibinfo{author}{\bibfnamefont{K.}~\bibnamefont{Kato}},
  \bibinfo{author}{\bibfnamefont{M.}~\bibnamefont{Mizumaki}},
  \bibinfo{author}{\bibfnamefont{T.}~\bibnamefont{Asaka}},
  \bibinfo{author}{\bibfnamefont{M.}~\bibnamefont{Takata}},
  \bibinfo{author}{\bibfnamefont{Y.}~\bibnamefont{Matsui}},
  \bibinfo{author}{\bibfnamefont{H.}~\bibnamefont{Sawa}}, \bibnamefont{and}
  \bibinfo{author}{\bibfnamefont{J.}~\bibnamefont{Akimitsu}},
  \bibinfo{journal}{Sci. Tech. Adv. Mat.} \textbf{\bibinfo{volume}{7}},
  \bibinfo{pages}{17} (\bibinfo{year}{2006}).

\bibitem[{\citenamefont{Baker et~al.}(2007)\citenamefont{Baker, Blundell,
  Pratt, Lancaster, Brooks, Hayes, Isobe, Ueda, Hoinkis, Sing
  et~al.}}]{bak2007}
\bibinfo{author}{\bibfnamefont{P.~J.} \bibnamefont{Baker}},
  \bibinfo{author}{\bibfnamefont{S.~J.} \bibnamefont{Blundell}},
  \bibinfo{author}{\bibfnamefont{F.~L.} \bibnamefont{Pratt}},
  \bibinfo{author}{\bibfnamefont{T.}~\bibnamefont{Lancaster}},
  \bibinfo{author}{\bibfnamefont{M.~L.} \bibnamefont{Brooks}},
  \bibinfo{author}{\bibfnamefont{W.}~\bibnamefont{Hayes}},
  \bibinfo{author}{\bibfnamefont{M.}~\bibnamefont{Isobe}},
  \bibinfo{author}{\bibfnamefont{Y.}~\bibnamefont{Ueda}},
  \bibinfo{author}{\bibfnamefont{M.}~\bibnamefont{Hoinkis}},
  \bibinfo{author}{\bibfnamefont{M.}~\bibnamefont{Sing}}, \bibnamefont{et~al.},
  \bibinfo{journal}{Phys. Rev. B} \textbf{\bibinfo{volume}{75}},
  \bibinfo{pages}{094404} (\bibinfo{year}{2007}).

\bibitem[{\citenamefont{Macovez}(2007)}]{Roberto}
\bibinfo{author}{\bibfnamefont{R.}~\bibnamefont{Macovez}}
  (\bibinfo{year}{2007}), \bibinfo{note}{unpublished}.

\bibitem[{\citenamefont{Noodleman and Norman}(1979)}]{Nood79}
\bibinfo{author}{\bibfnamefont{L.}~\bibnamefont{Noodleman}} \bibnamefont{and}
  \bibinfo{author}{\bibfnamefont{J.~G.} \bibnamefont{Norman}},
  \bibinfo{journal}{J. Chem. Phys.} \textbf{\bibinfo{volume}{70}},
  \bibinfo{pages}{4903} (\bibinfo{year}{1979}).

\bibitem[{\citenamefont{et~al.}()}]{Frish04}
\bibinfo{author}{\bibfnamefont{M.~J.~F.} \bibnamefont{et~al.}},
  \emph{\bibinfo{title}{Gaussian 03, revision c.02}}, \bibinfo{note}{gaussian,
  Inc., Wallingford, CT, 2004}.

\bibitem[{\citenamefont{Becke}(1993)}]{Beck93}
\bibinfo{author}{\bibfnamefont{A.~D.} \bibnamefont{Becke}},
  \bibinfo{journal}{J. Chem. Phys.} \textbf{\bibinfo{volume}{98}},
  \bibinfo{pages}{5648} (\bibinfo{year}{1993}).

\bibitem[{\citenamefont{R$\ddot{\textrm{u}}$ckamp
  et~al.}(2005{\natexlab{b}})\citenamefont{R$\ddot{\textrm{u}}$ckamp,
  Benckiser, Haverkort, Roth, Lorenz, Freimuth, Jongen,
  M$\ddot{\textrm{o}}$ller, Meyer, Reutler et~al.}}]{ruck2005long}
\bibinfo{author}{\bibfnamefont{R.}~\bibnamefont{R$\ddot{\textrm{u}}$ckamp}},
  \bibinfo{author}{\bibfnamefont{E.}~\bibnamefont{Benckiser}},
  \bibinfo{author}{\bibfnamefont{M.~W.} \bibnamefont{Haverkort}},
  \bibinfo{author}{\bibfnamefont{H.}~\bibnamefont{Roth}},
  \bibinfo{author}{\bibfnamefont{T.}~\bibnamefont{Lorenz}},
  \bibinfo{author}{\bibfnamefont{A.}~\bibnamefont{Freimuth}},
  \bibinfo{author}{\bibfnamefont{L.}~\bibnamefont{Jongen}},
  \bibinfo{author}{\bibfnamefont{A.}~\bibnamefont{M$\ddot{\textrm{o}}$ller}},
  \bibinfo{author}{\bibfnamefont{G.}~\bibnamefont{Meyer}},
  \bibinfo{author}{\bibfnamefont{P.}~\bibnamefont{Reutler}},
  \bibnamefont{et~al.}, \bibinfo{journal}{New Journal of Physics}
  \textbf{\bibinfo{volume}{7}}, \bibinfo{pages}{1367}
  (\bibinfo{year}{2005}{\natexlab{b}}).

\bibitem[{\citenamefont{Zakharov et~al.}(2006)\citenamefont{Zakharov,
  Deisenhofer, von Nidda, Lunkenheimer, Hemberger, Hoinkis, Klemm, Sing,
  Claessen, Eremin et~al.}}]{Zach2006}
\bibinfo{author}{\bibfnamefont{D.~V.} \bibnamefont{Zakharov}},
  \bibinfo{author}{\bibfnamefont{J.}~\bibnamefont{Deisenhofer}},
  \bibinfo{author}{\bibfnamefont{H.~A.~K.} \bibnamefont{von Nidda}},
  \bibinfo{author}{\bibfnamefont{P.}~\bibnamefont{Lunkenheimer}},
  \bibinfo{author}{\bibfnamefont{J.}~\bibnamefont{Hemberger}},
  \bibinfo{author}{\bibfnamefont{M.}~\bibnamefont{Hoinkis}},
  \bibinfo{author}{\bibfnamefont{M.}~\bibnamefont{Klemm}},
  \bibinfo{author}{\bibfnamefont{M.}~\bibnamefont{Sing}},
  \bibinfo{author}{\bibfnamefont{R.}~\bibnamefont{Claessen}},
  \bibinfo{author}{\bibfnamefont{M.~V.} \bibnamefont{Eremin}},
  \bibnamefont{et~al.}, \bibinfo{journal}{Phys. Rev. B}
  \textbf{\bibinfo{volume}{73}}, \bibinfo{pages}{094452}
  (\bibinfo{year}{2006}).

\bibitem[{\citenamefont{Karlstro et~al.}(2003)\citenamefont{Karlstro, Lindh,
  Malmqvist, Roos, Ryde, Veryazov, Widmark, Cossi, Schimmelpfennig, Neogrady
  et~al.}}]{Karl2003}
\bibinfo{author}{\bibfnamefont{G.}~\bibnamefont{Karlstro}},
  \bibinfo{author}{\bibfnamefont{R.}~\bibnamefont{Lindh}},
  \bibinfo{author}{\bibfnamefont{P.}~\bibnamefont{Malmqvist}},
  \bibinfo{author}{\bibfnamefont{B.}~\bibnamefont{Roos}},
  \bibinfo{author}{\bibfnamefont{U.}~\bibnamefont{Ryde}},
  \bibinfo{author}{\bibfnamefont{V.}~\bibnamefont{Veryazov}},
  \bibinfo{author}{\bibfnamefont{P.}~\bibnamefont{Widmark}},
  \bibinfo{author}{\bibfnamefont{M.}~\bibnamefont{Cossi}},
  \bibinfo{author}{\bibfnamefont{B.}~\bibnamefont{Schimmelpfennig}},
  \bibinfo{author}{\bibfnamefont{P.}~\bibnamefont{Neogrady}},
  \bibnamefont{et~al.}, \bibinfo{journal}{Comput. Mater. Sci.}
  \textbf{\bibinfo{volume}{28}}, \bibinfo{pages}{222} (\bibinfo{year}{2003}).

\bibitem[{\citenamefont{Wiedenmann et~al.}(1984)\citenamefont{Wiedenmann,
  Mignod, Venien, and Palvadeau}}]{wied84}
\bibinfo{author}{\bibfnamefont{A.}~\bibnamefont{Wiedenmann}},
  \bibinfo{author}{\bibfnamefont{J.~R.} \bibnamefont{Mignod}},
  \bibinfo{author}{\bibfnamefont{J.~P.} \bibnamefont{Venien}},
  \bibnamefont{and}
  \bibinfo{author}{\bibfnamefont{P.}~\bibnamefont{Palvadeau}},
  \bibinfo{journal}{JMMM} \textbf{\bibinfo{volume}{45}}, \bibinfo{pages}{275}
  (\bibinfo{year}{1984}).

\bibitem[{\citenamefont{Eckold}(1992)}]{esc92}
\bibinfo{author}{\bibfnamefont{G.}~\bibnamefont{Eckold}},
  \emph{\bibinfo{title}{UNISOFT - A Program Package for Lattice Dynamical
  Calculations: Users Manual}} (\bibinfo{year}{1992}).

\end{thebibliography}

\end{document}